\documentclass[11pt]{article} 

\usepackage{amsmath,amsthm,latexsym,amssymb,amsfonts,epsfig}

\newcommand{\be}{\begin{equation}}
\newcommand{\ee}{\end{equation}}
\newcommand{\ba}{\begin{eqnarray}}
\newcommand{\ea}{\end{eqnarray}}

\title{{\sf Asymptotically flat boundary conditions for}\\
 {\sf the $U(1)^3$ model for Euclidean Quantum Gravity}} 
\author{
{\sf S. Bakhoda}$^{1,2}$\thanks{{\sf 
s\_bakhoda@sbu.ac.ir, sepideh.bakhoda@gravity.fau.de}},
{\sf H. Shojaie}$^1$\thanks{{\sf
h-shojaie@sbu.ac.ir}},
{\sf T. Thiemann}$^2$\thanks{{\sf 
thomas.thiemann@gravity.fau.de}}\\
\\
{\sf $^1$ Dept. of Physics, Shahid Beheshti University,}\\
{\sf  G.C., Evin, Tehran 1983969411, Iran}\\
{\sf $^2$ Inst. for Quantum Gravity, FAU Erlangen -- N\"urnberg,}\\
{\sf Staudtstr. 7, 91058 Erlangen, Germany}\\
}
\date{{\small\sf \today}}

\makeatletter
\@addtoreset{equation}{section}
\makeatother

\begin{document} 

\maketitle

{\sf

\begin{abstract}
A generally covariant $U(1)^3$ gauge theory describing the $G_N \to 0$ limit of Euclidean general relativity is an interesting test laboratory for general relativity, specially because the algebra of the Hamiltonian and diffeomorphism constraints of this limit is isomorphic to the algebra of the corresponding constraints in general relativity. In the present work, we study boundary conditions and asymptotic symmetries of the $U(1)^3$ model and show that while asymptotic spacetime translations admit well-defined generators, boosts and rotations do not. Comparing with Euclidean general relativity, one finds that exactly the non-Abelian part of the $SU(2)$ Gauss constraint which is absent in the $U(1)^3$ model plays a crucial role in obtaining boost and rotation generators. 
\end{abstract}




\section{Introduction}
In the framework of the Ashtekar variables in terms of which General Relativity (GR) is formulated as a $SU(2)$ gauge theory \cite{Ashtekar}, attempts to find an operator corresponding to the Hamiltonian constraint
of Lorentzian vacuum canonical GR led to this intriguing result that the Lorentzian Hamiltonian constraint can be written as a sum of the Hamiltonian constraint of Euclidean vacuum GR and another term \cite{QSD1} which 
is under much control in the context of Loop Quantum Gravity \cite{Thomas}. This means that quantizing the Euclidean Hamiltonian constraint immediately gives us a quantized operator for the Lorentzian one \cite{QSD1}.
Although the idea followed in \cite{QSD1} has yielded significant results, 
the quantum constraint algebra, while closing, does so with the wrong 
structure ``constants'' and in that sense 
suffers from an anomaly.

Motivated to find a clue pushing the idea toward problem solving, one can study simpler theories and see what lessons can be learned from the outcome. The $G_N \to 0$ limit of the Euclidean gravity introduced by Smolin \cite{Smolin} is one of these models which is described by $U(1)^3$ gauge theory. 
This model contains three Gauss constraints, three spatial diffeomorphism constraints and a Hamiltonian constraint whose
constraint algebra for the Hamiltonian and diffeomorphism constraints is isomorphic to that of general relativity. This property in addition to the Abelian characteristic of its gauge group make the $U(1)^3$ theory an interesting toy model to scrutinize \cite{ Varadarajan}.
 
There exist two different approaches to work out the quantum theory for 
constrained theories. The first one, known as Dirac quantization 
\cite{Dirac}, quantizes the entire kinematical phase space producing a 
kinematical Hilbert space. Then physical states are those which are 
annihilated by all constraints operators acting on the kinematical Hilbert space. Therefore the physical sector of the theory in this approach is constructed in the quantum theory.
In the the second approach called reduced phase space quantization, one 
solves the constraints at the classical level and obtains 
a physical phase space whose variables are called observables which are 
gauge invariant quantities and then quantizes the physical phase space 
yielding to the physical Hilbert space. 
For the $U(1)^3$ model, there has been much recent progress
\cite{Varadarajan} and working on the latter has begun in \cite{B.T} 
where the analysis is restricted to spatial topology of 
$\mathbb{R}^3$ with the asymptotically flat boundary 
conditions. As asymptotically flat spacetimes are of great 
importance in GR, this paper is devoted to investigate their 
properties in $U(1)^3$ model and the results of the present paper
were used  
in \cite{B.T} which was in fact the main motivation for the present study.

To achieve asymptotic symmetry generators \cite{13, 14}, 
we seek for boundary terms to the constraints that produce well-defined 
phase space functions and Poisson brackets
while lapse function and shift vector obey decay behaviours 
corresponding to asymptotic symmetry transformations. 
When we are working in the context of Lorenzian or Euclidean GR, 
one expects these well-defined functions to generate the 
Poincar\'e or ISO(4) group respectively depending on 
the signature.
Regarding $U(1)^3$ theory, we examine to what extent we can recover 
ISO(4) transformations. In fact, the question is whether there are 
well-defined generators for all generators of ISO(4) in this model or not, 
and what the main reason for answering yes\slash no to this question is.
\\
\\
The architecture of this paper is as follows:
\\
\\
In section 2, we briefly review the background needed for the arguments 
following later. In subsection 2.1, first we express the constraints of 
GR in terms of both ADM  and $SU(2)$ variables. 
Then observing that they are not well-defined and functionally 
differentible, we revisit the results and reasoning, 
presented in \cite{Beig} and in 
\cite{Thiemann, Campiglia}, to improve the constraints to 
well-defined generators of Poincar\' e and ISO(4) group for Lorentzian 
and Euclidean GR respectively. 
In subsection 2.2, the $U(1)^3$ model will be concisely introduced 
and its constraints, 
on the basis of which the rest of the paper is written, are expressed.

In section 3, we try to make the constraints well-defined. 
To do this, first we look for a boundary term destroying 
differentiability of the constraint and then subtract it from the variation 
of the constraint. If the boundary term is an exact one form 
in the field space, it turns out that this expression is functionally differentiable. At this point, there is only one step left for the constraint to be well-defined, and that is to examine its finiteness. 
we show that in the $U(1)^3$ model, while spacetime translations 
are allowed asymptotic symmetries, 
there are no well-defined generators for boosts and rotations.  

In section 4, we compare the results of sections 2.1 and 3 and 
exhibit the main source of ill-definedness of boost and rotation 
generators in the $U(1)^3$ model. 

In the last section we conclude with a brief summary.
\section{Background}

\subsection{Review of asymptotically flat boundary conditions for the $SU(2)$ case}
In general, consistent boundary conditions are supposed to provide a well-defined symplectic structure and the finite and integrable charges associated with the asymptotic symmetries. Here integrability means that the variation of the surface charge is an exact one form.
In the ADM formulation of asymptotically
flat spacetimes, it is assumed that on spatial slices asymptotic spheres are equipped with asymptotically cartesian coordinates $x^a$ at spatial infinity; i.e. $r \to \infty$ where $r^2 = x^ax_a$. 
Taking this as the starting point and seeking for an appropriate boundary conditions, one
sees that on any hypersurface, the fall-off behaviours of the spatial metric $q_{ab}$ and its conjugate momentum $\pi^
{ab}$ have to be
\begin{equation}\label{ADM BC}
    \begin{split}
        q_{ab}=& \delta_{ab}+\frac{h_{ab}}{r}+\mathcal{O}(r^{-2})\\
        \pi^{ab}=&\frac{p^{ab}}{r^2}+\mathcal{O}(r^{-3})
    \end{split}
\end{equation}
where $h_{ab}$ and $p^{ab}$ are smooth tensor fields on the asymptotic 2-sphere. In (\ref{ADM BC})
the first condition follows directly from the form of the spacetime metric for the asymptotically flat case, while the second one is a consequence of demanding a non-vanishing ADM momentum.
In order to eliminate the logarithmic singularity existing in the symplectic structure, the leading terms in (\ref{ADM BC}) need to admit additional certain parity conditions as 
\begin{equation}\label{ADM PC}
    \begin{split}
        h_{ab}\left(-\frac{x}{r}\right)=h_{ab}\left(\frac{x}{r}\right), \; \; \; p^{ab}\left(-\frac{x}{r}\right)=-p^{ab}\left(\frac{x}{r}\right)
    \end{split}
\end{equation}

Indeed, the integral of $p^{ab}\Dot{h}_{ab}$ over the sphere, which is the coefficient of the
singularity, vanishes owing to (\ref{ADM PC}).
The parity conditions also
give rise to the finite and integrable Poincar\'e or ISO(4) charges.

Furthermore, aiming to retain the boundary conditions (\ref{ADM BC}) invariant under the hypersurface deformations, 
\begin{equation}\label{HD}
    \begin{split}
        \delta q_{ab} = & \frac{-2sN}{\sqrt{q}}(\pi_{ab} - \frac{1}{2}\pi q_{ab}) + \mathcal{L}_{\vec{N}}q_{ab}\\
        \delta \pi^{ab} = & -N\sqrt{q}({^{(3)}R}^{ab} - \frac{1}{2}q^{ab}{^{(3)}R}) - \frac{sN}{2\sqrt{q}}(\pi_{cd}\pi^{cd} - \frac{1}{2}\pi^2)q^{ab} \\
        & + \frac{2sN}{\sqrt{q}}(\pi^{ac}{\pi_c}^b - \frac{1}{2}\pi^{ab}\pi) + \sqrt{q}(D^aD^b N - q^{ab}D_a D^b N)
         + \mathcal{L}_{\Vec{N}}\pi^{ab},
    \end{split}
\end{equation}
one requires to restrict the fall-off behaviours of the lapse function, $N$, and the shift vector, $N^a$. 
In (\ref{HD}), $q:=\det (q_{ab})$, ${^{(3)}R}_{ab}$ is the Ricci tensor of the spatial hypersurface, $D_a$ is the torsion free, metric compatible connection with respect to $q_{ab}$ and $s$ denotes the signature of spacetime mertic, i.e. $s = +1$ and $s=-1$ for Euclidean and Lorentzian spacetimes respectively. It turns out that the most general behaviour of them which also includes the generators of the asymptotic Poincar\'e and ISO(4) groups are 
 \begin{equation}\label{L&SH-Cartesian}
    \begin{split}
        & N = \beta_ax^a + \alpha + S +\mathcal{O}(r^{-1})\\
        & N^a = {\beta^a}_bx^b + \alpha^a + S^a + \mathcal{O}(r^{-1}),
    \end{split}
\end{equation}
where $\beta_a$ and $\beta_{ab}(= -\beta_{ba})$ are arbitrary constants representing boosts and rotations. Here, if $v^a$ denotes the velocity, then the boost parameter is $\beta^a= \frac{v^a}{\sqrt{1+sv^2}}=:\gamma v^a$ which satisfies the identity $\gamma^2 + s \gamma^2 v^2 =1$. This identity indicates that for a Euclidean boost, when $s=1$, the sine and cosine appear in the transformation matrix which says that the Euclidean boost is nothing but a rotation in the $x^0, \vec{x}$ plane. In turn, the arbitrary function $\alpha$, and arbitrary vector $\alpha^a$ represent time and spatial translations respectively and $S$ , $S^a$ which are odd functions on the asymptotic $S^2$ correspond to supertranslations.

On the other hand, since in vacuum GR the Hamiltonian and diffeomorphism constraints, i.e. 

\begin{equation}\label{S-C}
     \begin{split}
        &H[N]:=\int d^3x \; N\left(\frac{-s}{\sqrt{q}}\left[(q_{ac}q_{bd}-\frac{1}{2}q_{ab}q_{cd})\pi^{ab}\pi^{cd}\right]-\sqrt{q} \;{^{(3)}R}\right) 
        \\
        &H_a[N^a]:=-2\int d^3x \; N^a D_b \pi^b_a,
     \end{split}
 \end{equation}
are the generators of gauge transformations, they have to be finite and functionally differentiable so that their poisson bracket with any function on the phase space can be computed.
With regard to (\ref{ADM BC}) and (\ref{L&SH-Cartesian}), it is easy to check that the constraints (\ref{S-C}) are neither finite nor differentiable. To remedy this situation, a surface integral spoiling differentiability should be subtracted from the variation of the  constraint functionals. As mentioned before, appropriate boundary conditions result in an exact surface term and thus one can define new expressions for the constraints which now are functionally differentiable. Last step is to examine the convergence of these expressions. Having done this procedure, the authors of \cite{Beig} obtained the following well-defined constraints 
\begin{equation}
    \begin{split}
        &J[N]:= H[N]+ 2\oint dS_d\; \sqrt{q}q^{a[b} q^{c]d}[N \partial_b q_{ca}-\partial_b N(q_{ca}-\delta_{ca})]\\
        &J_a[N^a]:= H_a[N^a] + 2 \oint dS_a \; N_b \pi^{ab}
    \end{split}
\end{equation}
where $\oint$ is the integration over the asymptotic 2-sphere. 
\\

This analysis in terms of ADM variables language can be translated to 
the Ashtekar variables $(A_a^i, E^a_i)$ where the connection, $A^i_a$, is an $su(2)$-valued one form and its momentum conjugate, $E^a_i$, is a densitized 3-Bein. However, achieving this is challenging since there is an internal $su(2)$ frame whose asymptotic behaviour has to be determined. 

Accordingly, the boundary conditions (\ref{ADM BC}) and (\ref{ADM PC}) in terms of the Ashtekar variables can be written as
\begin{equation}\label{Ashtekar BC}
    \begin{split}
        &E^a_i=\delta^a_i+\frac{f^a_i}{r}+\mathcal{O}(r^{-2})\\
        &A_a^i=\frac{g^i_a}{r^2}+\mathcal{O}(r^{-3})
    \end{split}
\end{equation}
where
\[   
\delta^a_i= 
     \begin{cases}
       1 &\quad\text{if}\; (a,i)=(x,1), (y,2), (z,3),\\
       0 &\quad\text{otherwise.} \\ 
     \end{cases}
\]
and $f^a_i$ and $g_a^i$ are tensor fields defined on the asymptotic 2-sphere with the following definite parity conditions
\begin{equation}\label{Ashtekar PC}
    \begin{split}
        f^a_i\left(-\frac{x}{r}\right)=f^a_i\left(\frac{x}{r}\right), \; \; \; g^i_a\left(-\frac{x}{r}\right)=-g^i_a\left(\frac{x}{r}\right).
    \end{split}
\end{equation}
By the decay conditions (\ref{Ashtekar BC}) and (\ref{Ashtekar PC}), it is assured that the symplectic structure is well-defined.

In Euclidean GR, the constraint surface in the $(A,E)$-phase space is 
defined by the vanishing 
of the following functionals called Gauss, Hamiltonian and diffeomorphism constraints respectively

\begin{equation}\label{Ashtekar Constraints}
    \begin{split}
        &G_i[\Lambda^i]=\int d^3x\; \Lambda^i \left(\partial_a E^a_i+\epsilon_{ijk}A^j_a E^a_k\right)\\
        &H[N]= \int d^3x\; N \epsilon_{ijk}F^i_{ab}E^a_jE^b_k \\
        &H_a[N^a]=\int d^3x \; N^a\left(F^j_{ab}E^b_j-A^j_aG_j\right).
    \end{split}
\end{equation}
\\
Here
\begin{equation}\label{Fab}
    F^i_{ab}=\partial_aA^i_b-\partial_bA^i_a+{\epsilon^i}_{jk}A^j_aA^k_b ,
\end{equation}
$\Lambda^i$ is the Lagrange multiplier associated with ${G}_i$ and $N$ is the densitized lapse function with weight $-1$. It is desired to attain well-defined form of these functionals with the smearing functions including ISO(4) generators (\ref{L&SH-Cartesian}). To do this, first one has to ascertain an appropriate decay behaviour for $\Lambda^i$. Since the leading term of $G_i$ is $\mathcal{O}(r^{-2})$ odd, convergence of $G_i[\Lambda^i]$ requires the following fall-off condition
\begin{equation}\label{BC for lambda}
    \begin{split}
        \Lambda^i = \frac{\lambda^i}{r}+\mathcal{O}(r^{-2})
    \end{split}
\end{equation}
where $\lambda^i$ are even functions defined on the asymptotic 2-sphere.
It is straightforward to verify that (\ref{BC for lambda}) also ensures the differentiability of $G_i[\Lambda^i]$. Even after subtracting the surface integral destroying differentiability of the Hamiltonian and diffeomorphism constraints (\ref{Ashtekar Constraints}), it turns out that they are convergent only for  translations and not for boosts and rotations. This situation should be cured in a way that 1) the generators stay functionally differentiable and 2) the well-defined generator for translations which is already available can be recovered up to a pure gauge.
As shown in \cite{Thiemann, Campiglia}, ultimately the well-defined forms of the symmetry generators are
\begin{equation}\label{final ex}
    \begin{split}
        &J[N]=H[N]-\oint dS_a \; N \epsilon_{ijk}A^i_b E^a_j E^b_k-G_i[\Lambda^i_B]+ \oint dS_a\; E^a_i \Bar{\Lambda}^i_B\\
        &J_a[N^a]=H_a[N^a]-\oint dS_a\; N^a A^i_b E^b_i-G_i[\Lambda^i_R]
        + \oint dS_a\; E^a_i \Bar{\Lambda}^i_R
    \end{split}
\end{equation}
where $\Lambda^i_R= \Lambda^i + \Bar{\Lambda}^i_R= \Lambda^i -\frac{1}{2}\epsilon_{ijk}\delta^j_a \delta^b_k \beta^a_b$ and $\Lambda^i_B= \Lambda^i + \Bar{\Lambda}^i_B= \Lambda^i +\delta^a_i \beta_a$. The second term appearing in either expressions in (\ref{final ex}) is the surface term subtracted to make the original functionals (\ref{Ashtekar Constraints}) differentiable. The third term is subtracted to get rid of the source of divergence for boosts and rotations but puts them again in the status of non-differentiability which is modified by adding the last term. As expected, the volume terms added to the constraints are proportional to the Gauss constraint and thus do not change the translation generator on the constraint surface of the Gauss constraint.

\subsection{Review of $U(1)^3$ model for Euclidean quantum gravity}
In~\cite{Smolin}, Smolin introduced the weak field limit of the Euclidean gravity $G_N\rightarrow0$, where $G_N$ is the Newtonian gravitational constant, by expanding the phase space variables $(A,E)$ as
\begin{equation}\label{Ashtekar}
    \begin{split}
    E=E_0+G_N E_1+G^2_N E_2+...\\
    A=A_0+G_N A_1+G^2_N A_2+... ,
    \end{split}
\end{equation}
at the level of the action. The resulting theory
is not to be confused with standard perturbation theory.
More precisely, consider the Hamiltonian for Euclidean gravity
\begin{equation}\label{Hamil}
         \mathcal{H}[E,A]=\frac{1}{G_N}\int d^3x\left( N^aH_a+NH+\Lambda^iG_i\right)
\end{equation}
Rescaling the dimensionful quantities in~(\ref{Hamil}) by $G_N$, namely the connection $A^i_a\rightarrow G_NA^i_a$ and the Lagrange multiplier $\Lambda^i\rightarrow G_N\Lambda^i$, the Gauss constraint of ~(\ref{Ashtekar Constraints}) and~(\ref{Fab}) become 
\begin{equation}\label{FandG}
    \begin{split}
       G_i=&\mathcal{D}_aE^a_i=\partial_aE^a_i+{\epsilon_{ij}}^kG_NA^j_aE^a_k \\
       F^i_{ab}=&\partial_aA^i_b-\partial_bA^i_a+{\epsilon^i}_{jk}G_NA^j_aA^k_b .
    \end{split}
\end{equation}
respectively.  From~(\ref{FandG}), it is obvious that the internal gauge symmetry is still $SU(2)$. However, in the limit $G_N\rightarrow0$, the second term which brings in the self-interaction is switched off. The Poisson bracket of a pair of Gauss constraints then commutes as one has
\begin{equation}
		\{G_i[\Lambda^i],G_j[\Delta^j]\}\propto G_N ,
\end{equation} 
and the symmetry group contracts from $SU(2)$ to three  independent Abelian internal gauge symmetry $U(1)$, namely $U(1)^3$, each of which corresponds to one of the gauge fields $A^i$ ($i=1,2,3)$. Consequently, the constraints remain first class and have the following simpler forms
\begin{equation}\label{Cons.S.}
    \begin{split}
        &C_j[\lambda^j]=\int d^3x\; \Lambda^j \partial_a E^a_j\\
        &C_a[N^a]=\int d^3x\; N^a\left(F^j_{ab}E^b_j-A^j_a\partial_bE^b_j\right)\\
        &C[N]=\int d^3x\; N F^j_{ab}E^a_k E^b_l \epsilon_{jkl}
    \end{split}
\end{equation}
where $C_j$, $C_a$ and $C$ are the Gauss, diffeomorphism and Hamiltonian constraints for the $U(1)^3$ model respectively and $F^j_{ab}=\partial_a A^j_b - \partial_b A^j_a$ is the corresponding curvature. The Hamiltonian then reads as
\begin{equation}\label{Hamil}
         \mathcal{H}[E,A]=\frac{1}{G_N}\int d^3x\left( N^aC_a+N C+\Lambda^iC_i\right)
\end{equation}
and the only non-vanishing Poisson brackets of the pair of the constraints in the algebra will be
\begin{equation}
    \begin{split}
    \{C_a[N^a],C_b[M^b]\}&=C_a[\mathcal{L}_{\overrightarrow{N}} M^a]\\
    \{C_a[N^a],C[N]\}&=C[\mathcal{L}_{\overrightarrow{N}}N] \\ 
    \{C[N],C[M]\}&=C_a[E^{ia}E^b_i(N\mathcal{D}_b M-M\mathcal{D}_b N)].
    \end{split}
\end{equation}
\\
The algebra, except for the vanishing Poisson bracket of a pair of Gauss constraints, $C_i$'s, is isomorphic to the algebra of GR as it can be easily seen. To see some work done in this context, we refer the reader to \cite{Varadarajan}.

\section{Generators of Asymptotic symmetries for $U(1)^3$ model}
As the $U(1)^3$ model is a test laboratory for GR, it is of interest to know whether boundary conditions and asymptotic symmetries of these two theories are identical. The question to be answered in this section is whether the ISO(4) group can be considered as the asymptotic symmetries of the model. In other words, is there a way to construct well-defined functionals from the constraints (\ref{Cons.S.}) while the lapse and the shift include the ISO(4) generators? Although, the model being persued is not Euclidean GR and we don't presume that the ISO(4) group is the asymptotic symmetry, it is appealing to investigate to what extent the model admits a subgroup of the ISO(4) group. In what follows we examine the constraints (\ref{Cons.S.}) and try to make them well-defined.

\subsection*{Gauss constraint}
The action of the Gauss constraint on the phase space variables is
\begin{equation}
    \begin{split}
        &\delta_{\Lambda}A^j_a=\{C_i[\Lambda^i],A^j_a\}=-\partial_a\Lambda^j\\
        &
        \delta_{\Lambda}E^a_j=\{C_i[\Lambda^i],E^a_j\}=0
    \end{split}
\end{equation}
Thus one sees that
\begin{equation}
    \begin{split}
        \delta C_j[\Lambda^j]=&\int d^3x \; \Lambda^j \partial_a \delta E^a_j=\oint dS_a \; \Lambda^j \delta E^a_j
        -\int d^3x \; (\partial_a\Lambda^j) \delta E^a_j=-\int d^3x \; (\partial_a\Lambda^j) \delta E^a_j\\
        =&
        \int d^3x\; \left[(\delta_\Lambda A^j_a) \delta E^a_j - (\delta_\Lambda E^a_j) \delta A^j_a \right]
    \end{split}
\end{equation}
is functionally differentiable. Here the surface term has been dropped because $\delta E^a_j=\mathcal{O}(r^{-2})$ and $\Lambda^j = \mathcal{O}(r^{-1})$.  
As $\partial_a E^a_j= O(r^{-2})$ odd, the integrand of $C_j[\Lambda^j]$ is $\mathcal{O}(r^{-3})$ odd and hence the constraint is also finite.
\\

\subsection*{Scalar constraint}
It is straightforward to see that
\begin{equation}
    \begin{split}
        &\delta_N A_c^i (x)=\{C[N],A^i_c (x)\}
        =-2 N \epsilon_{ijk}F^j_{ac}E^a_k\\
        & 
        \delta_N E^c_i (x)=\{C[N],E^c_i (x)\}
        =2\epsilon_{ikl}\partial_b(NE^c_kE^b_l)
    \end{split}
\end{equation}
\\
Thus the variation of this constraint is

\begin{equation}\label{Variation of C}
    \begin{split}
        \delta C[N]&= \int d^3x \; \epsilon_{jkl}N \left(\delta F^j_{ab}E^a_kE^b_l+2 F^j_{ab}E^b_l \delta E^a_k \right)\\
        &= \int d^3x \; \epsilon_{jkl}N \left(( \partial_a \delta A^j_{b}-\partial_b \delta A^j_{a})E^a_kE^b_l+2 F^j_{ab}E^b_l \delta E^a_k \right)\\
        &=2\int d^3x \; \epsilon_{jkl} \left(\partial_a(NE^a_k E^b_l \delta A^j_b)-\delta A^j_{b}\partial_a(N E^a_kE^b_l)+ NF^j_{ab}E^b_l \delta E^a_k \right)\\
        &=\int d^3x \; 2\epsilon_{jkl} \left(NF^j_{ab}E^b_l \delta E^a_k-\delta A^j_{b}\partial_a(N E^a_kE^b_l)\right)+2\int dS_a(NE^a_k E^b_l \delta A^j_b)\\
        &=\int d^3x \; \left(\delta A^j_{b}(\delta_N E^b_j)-(\delta_N A^k_a) \delta E^a_k\right)+2\delta \oint dS_a\; \epsilon_{jkl}(NE^a_k E^b_l  A^j_b)\\
    \end{split}
\end{equation}
\\
We have pulled the variation out of the surface
integral in (\ref{Variation of C}) because the correction terms are $\mathcal{O}(r^{-1})$ even for a translation and $\mathcal{O}(1)$ odd for a boost. Now we define the new generator as
\begin{equation}\label{New scalar generator}
    \begin{split}
        C'[N]:= C[N]-2\oint dS_a\; N \epsilon_{jkl} A^j_b E^a_k E^b_l 
    \end{split}
\end{equation}
which is functionally differentiable. At this step one is supposed to check whether it is finite.
\begin{equation}
    \begin{split}
        C'[N]=&\int d^3x \; \epsilon_{jkl} \left(F^j_{ab}E^a_k E^b_l N - 2\partial_a (A^j_b E^a_k E^b_l N) \right)\\
        =&2\int d^3x \; \epsilon_{jkl} \left(\frac{1}{2}F^j_{ab}E^a_k E^b_l N -  E^a_k E^b_l N\partial_a A^j_b- A^j_b E^b_l N\partial_a E^a_k- A^j_b E^a_k N\partial_a E^b_l - A^j_b E^a_k E^b_l \partial_aN \right)\\
        =&-2\int d^3x \; \epsilon_{jkl} \left(A^j_b E^b_l N\partial_a E^a_k+ A^j_b E^a_k N\partial_a E^b_l + A^j_b E^a_k E^b_l \partial_aN \right)\\
    \end{split}
\end{equation}
Here terms of the form $AEN \partial E$ are $\mathcal{O}(r^{-4})$ even for a translation and $\mathcal{O}(r^{-3})$ odd for a boost. Thus they are convergent. The last term which is of the form $AEE\partial N$ vanishes for a translation but is divergent for a boost. Therefore, $C'[N]$ is well-defined for a translation and the source of its divergence for a boost is 
\begin{equation}\label{source of Div. boost}
     \begin{split}
        -2\int d^3x\; \epsilon_{jkl} (A^j_b E^a_k E^b_l \beta_a) &=- \int d^3x\; \frac{1}{r^2}(\beta_a \epsilon_{jkl}g^j_b \delta^a_k \delta^b_l)-
        \int d^3x\; \beta_a \epsilon_{jkl}A^j_b \delta^a_k E^b_l+ \text{finite}\\
        &=-
        \int d^3x\; \beta_a \epsilon_{jkl}A^j_b \delta^a_k E^b_l+ \text{finite}
    \end{split}
\end{equation}
\\
where in going from the first line to the second one we have used the parity of $g^j_b$ to drop the linear singularity and so we are left with the logarithmic singularity.
Wishing to get rid of the divergence, one has to subtract (\ref{source of Div. boost}) from  (\ref{New scalar generator}) but this would play an undesirable role on the constraint surface since (\ref{source of Div. boost}) is proportional neither to the constraints nor to a part of them. Consequently, time translations have a well-defined generator (\ref{New scalar generator}), but boosts do not! 
A more detailed argument is given in section \ref{section 4}.

\subsection*{Vector constraint}
The vector constraint acts on the canonical variables as follows
\begin{equation}
    \begin{split}
        \delta_{\Vec{N}}A^i_c(x)
        =&\{C_a[N^a], A^i_c(x)\} =-\mathcal{L}_{\Vec{N}}A^i_c\\
        \delta_{\Vec{N}}E^c_i (x) =&\{C_a[N^a], E^c_i (x)\} =-\mathcal{L}_{\Vec{N}}E^c_i
    \end{split}
\end{equation}
So, variation of the constraint is
\begin{equation}
  \begin{split}
    \delta C_a[N^a]
    =&\int d^3x\;N^a\left(\delta  F^j_{ab}E^b_j+F^j_{ab}\delta E^b_j-\delta A^j_a\partial_bE^b_j-A^j_a\partial_b\delta E^b_j\right)\\
    =&\int d^3x\;N^a\left(\partial_a\delta A^j_bE^b_j-\partial_b\delta A^j_aE^b_j+\partial_a A^j_b\delta E^b_j-\partial_bA^j_a\delta E^b_j-\delta A^j_a\partial_bE^b_j-A^j_a\partial_b\delta E^b_j\right)\\
    =&\int d^3x\;\left(\delta A^j_a\partial_b (N^aE^b_j)-\delta A^j_b\partial_a(N^aE^b_j)+N^a\partial_a A^j_b\delta E^b_j-N^a\partial_b A^j_a\delta E^b_j\right. \\
    &\left.\hspace{1cm}-N^a\delta A^j_a\partial_b E^b_j+\partial_b(N^a A^j_a)\delta E^b_j\right)
    +\oint dS_a \; (N^a E^b_j \delta A^j_b - N^b E^a_j \delta A^j_b - N^bA^j_b\delta E^a_j )\\
    =&\int d^3x\;\left(\delta A^j_a\left[E^b_j\partial_b N^a-\partial_b(N^b E^a_j)\right]
    +\delta E^b_j \left[N^a\partial_a A^j_b+A^j_a\partial_b N^a \right]\right)\\
    &+\oint d S_a \; (N^a E^b_j \delta A^j_b - N^b E^a_j \delta A^j_b)\\
    =&\int d^3x\;\left(\delta A^j_a\left[-\mathcal{L}_{\vec{N}} E^a_j\right]
    +\delta E^b_j \left[\mathcal{L}_{\vec{N}} A_b^j \right]\right)
    +\oint d S_a \; (N^a E^b_j \delta A_j^b - N^b E^a_j \delta A^j_b)\\
    =&\int d^3x\;\left(\delta A^j_a\left[\delta_{\vec{N}} E^a_j\right]
    -\delta E^b_j \left[\delta_{\vec{N}} A_b^j \right]\right)
    +\delta \oint d S_a \; (N^a E^b_j  A^j_b - N^b E^a_j A^j_b)\\
  \end{split}
\end{equation}
Here the third term of the surface integral in the fourth line is $\mathcal{O}(1)$ odd for a rotation and  $O(r^{-1})$ even for a translation and so can be put away. Furthermore, in the last step one can pull the variation out of the surface integral since the correction terms are $\mathcal{O}(r^{-1})$ even for a translation and $\mathcal{O}(1)$ odd for a rotation.

So the new generator should be defined as
\begin{equation}\label{New vector generator}
    C'_a[N^a]:= C_a[N^a]-\oint dS_a\left(N^aE^b_j-N^bE^a_j\right)A^j_b
\end{equation}
which is functionally differentiable. To check its finitess, we rewrite (\ref{New vector generator}) as a volume integral 

\begin{equation}\label{finiteness of NG VC}
     \begin{split}
         C'_a[N^a]=&\int d^3x \; \left[N^aF^j_{ab}E^b_j-N^aA^j_a\partial_bE^b_j-\partial_a(N^aE^b_jA^j_b-N^bE^a_jA^j_b)\right]\\
         =&\int d^3x \; \left[N^aF^j_{ab}E^b_j-N^aA^j_a\partial_bE^b_j-N^aE^b_j F^j_{ab}-\partial_a(N^aE^b_j-N^bE^a_j)A^j_b\right]\\
         =&-\int d^3x \; \left[N^aA^j_a\partial_bE^b_j+\partial_a(N^aE^b_j-N^bE^a_j)A^j_b\right]\\
         =&-\int d^3x \; A^j_b\left[E^b_j\partial_a N^a+N^a\partial_a E^b_j -E^a_j \partial_a N^b\right]=-\int d^3x\; A^j_b\mathcal{L}_{\Vec{N}}E^b_j\\
      \end{split}
\end{equation}
\\
In the last line of (\ref{finiteness of NG VC}), the term $AN\partial E$ is $\mathcal{O}(r^{-4})$ even for a translation and $\mathcal{O}(r^{-3})$ odd for a rotation which means it is convergent. $A^j_b E^b_j \partial_a N^a$ vanishes for both translation and rotation because $\alpha^a$ is a constant and $\beta^a_b$ is antisymmetric. On the other hand, the other term of the form $A E\partial N$ is $\mathcal{O}(r^{-2})$ odd for a rotation and vanishes for a translation. Thus, $C'_a[N^a]$ is well-defined for a translation and the source of its divergence for a rotation is  
\begin{equation}\label{source of Div. rotation}
    \begin{split}
         \int d^3x \; A^j_b E^a_j \beta_a^b=&\int d^3x \; \beta_a^b A^j_b \left(\delta^a_j + \frac{f^a_j}{r}+\dots \right) \\
         =& \int d^3 x\; \beta_a^b A^j_b \delta^a_j + \int d^3 x\; \beta_a^b A^j_b \frac{f^a_j}{r} + \text{finite}\\
         =&  \int d^3 x\; \beta_a^b A^j_b \delta^a_j +\text{finite}
     \end{split}
\end{equation}
where the second integral in the second line is convergent since its integrand is $\mathcal{O}(r^{-3})$ odd. Again, (\ref{source of Div. rotation}) can not be written in terms of the constraints and thus subtracting the volume integral from (\ref{New vector generator}) is not allowed. 
Consequently, spatial translations have a well-defined generator (\ref{New vector generator}), but rotations do not! 

\section{Comparison with the $SU(2)$ case}\label{section 4}
In this section, we intend to place the situation under scrutiny and see what exactly causes the difference between $U(1)^3$ and $SU(2)$ cases that the former does not admit generators for boosts and rotations, but the latter does.
First, we split $F^i_{ab}$ and $G_i$ into its Abelian and non-Abelian parts, i.e. $F^i_{ab}=F^{i+}_{ab}+F^{i-}_{ab}$ and $G_i=G_i^+ + G_i^-$ where $F^{i+}_{ab}=\partial_a A^i_b-\partial_b A^i_a$,  $F^{i-}_{ab}=\epsilon_{ijk}A^j_aA^k_b$, $G_i^+=\partial_a E^a_i$ and $G_i^-=\epsilon_{ijk}A^j_a E^a_k$; accordingly the Hamiltonian and diffeomorphism constraints have also two parts corresponding to the plus and minus pieces of $F^i_{ab}$ and $G_i$, namely $H[N]=H^+[N]+H^-[N]$ and $H_a[N^a]=H_a^+[N^a]+H_a^-[N^a]$, where
\begin{align}
    H^+[N]=\int d^3x \; N\epsilon_{jkl} F^{j+}_{ab}E^a_k E^b_l &, \; \; \; \; H^-[N]=\int d^3x \; N\epsilon_{jkl} F^{j-}_{ab}E^a_k E^b_l \\
        H_a^+[N]=\int d^3x \; N^a(F^{j+}_{ab}E^b_j-A^j_a G_j^+) &, \; \; \; \; H_a^-[N]=\int d^3x \; N^a(F^{j-}_{ab}E^b_j-A^j_a G_j^-)=0.
\end{align}

Due to the boundary conditions, $F^{i-}_{ab}= \mathcal{O}(r^{-4})$ even and $G^-_i=\mathcal{O}(r^{-2})$ odd. Hence, the integrand of $H^-[N]$ is $\mathcal{O}(r^{-4})$ even for a translation and $\mathcal{O}(r^{-3})$ odd for a boost. Therefore the minus parts of these constraints are already finite. 
We show that $H^-[N]$ is also functionally differentiable. Its action on the canonical variables is   
\begin{equation}\label{3}
    \begin{split}
        &\delta_{N^-}A_c^l:=\{H^-[N], A_c^l (x)\}=-2N \epsilon_{ilk}\epsilon_{imn}A^m_c A^n_b E^b_k\\
        &\delta_{N^-}E^c_l:=\{H^-[N], E^c_l (x) \}=2N\epsilon_{ijk}\epsilon_{iln}E^c_j E^b_k A^n_b
    \end{split}
\end{equation}
Using (\ref{3}), one observes
\begin{equation}
    \begin{split}
        \delta H^-[N]&=\int d^3x \; N\epsilon_{jkl}(\delta F^{j-}_{ab}E^a_k E^b_l+F^{j-}_{ab}\delta E^a_k E^b_l+F^{j-}_{ab}E^a_k \delta E^b_l) \\
        &=\int d^3x \; (\delta A^l_c (2N\epsilon_{ijk}\epsilon_{iln}E^c_j E^b_k A^n_b) + \delta E^c_l (2N \epsilon_{ilk}\epsilon_{imn}A^m_c A^n_b E^b_k)) \\
        &=\int d^3x \; (\delta A^l_c (\delta_{N^-}E^c_l) - \delta E_l^c (\delta_{N^-}A_c^l)) 
    \end{split}
\end{equation}
\\
which means $H^-[N]$ is differentiable. Consequently, what need to be modified are $H^+[N]=C[N]$ and $H_a^+[N^a]=C_a[N^a]$, thus all failure to be well-defined is rooted in the $U(1)^3$ part of the Hamiltonian and diffeomorphism constraints existing in the $SU(2)$ case. So, as long as finding the source of divergence and non-differentiability is concerned calculations are the same in both cases. This brings us back to (\ref{source of Div. boost}) and (\ref{source of Div. rotation}) for boosts and rotations respectively.

For a boost the source of divergence can be expressed as
\begin{equation}\label{1}
    \begin{split}
    -\int d^3x\; \beta_a \epsilon_{jkl}A^j_b \delta^a_k E^b_l
    &=\int d^3x\; (\beta_a \delta^a_k) G^-_k 
    =\int d^3x\; \Bar{\Lambda}^k_B (G_k - G^+_k)\\
    &=G_k[\Bar{\Lambda}^k_B]-\int d^3x\;  \partial_b (\Bar{\Lambda}^k_B E^b_k)\\
    &=G_k[\Bar{\Lambda}^k_B]-\oint dS_b \Bar{\Lambda}^k_B E^b_k
    \end{split}
\end{equation}
\\
where we used $\partial_b \Bar{\Lambda}^k_B = 0$.  
As expected, at the end the volume term is proportional to $G_k$ and does not have any impact on constraint surface. One can verify that $J[N]$ in (\ref{final ex}) is the final well-defined generator.
Actually the subtle point is exactly here. In the $U(1)^3$ case, the absence of $G^-_k$ is responsible for excluding a well-defined boost generator. 

Going to (\ref{source of Div. rotation}) and trying to get rid of it, one sees

\begin{equation}
    \begin{split}
        \int d^3 x\; A^j_b \beta_a^b \delta^a_j &= \int d^3 x\; A^j_b (\epsilon_{ijk}\delta_k^b \Bar{\Lambda}^i_R)\\
        &=\int d^3x \; A^j_b (\epsilon_{ijk} E_k^b \Bar{\Lambda}^i_R)-\int d^3x \; A^j_b (\epsilon_{ijk} \frac{f_k^b}{r} \Bar{\Lambda}^i_R) +\text{finite}\\
        &=\int d^3x \; \Bar{\Lambda}^i_R G_i^- +\text{finite}\\
        &=\int d^3x \; \Bar{\Lambda}^i_R (G_i-G_i^+) +\text{finite}\\
        &=G_i[\Bar{\Lambda}^i_R]- \oint d S_a  \; \Bar{\Lambda}^i_R E_i^a +\text{finite}
    \end{split}
\end{equation}
where $\partial_a \Bar{\Lambda}^i_R=0$ has been used and the second integral in the second line is dropped since it is $\mathcal{O}(r^{-3})$ odd. Again the volume term is proportional to the Gauss constraint as desired. It is straightforward to investigate that $J_a[N^a]$ is the well-defined generator for the spatial translations and rotations. Here, just like as (\ref{1}), presence of $G_i^-$ (which is zero in the $U(1)^3$ case) plays a crucial role to obtain the generator. 

In addition to the technical reasoning one can add a further intuitive one:
The $SU(2)$ Gauss constraint generates rotations on the internal tangent space associated with the internal indices $j,k,l,\dots$ while asymptotic 
rotations act on the spatial tangent space corresponding to the indices $a,b,c,\dots$ . Due to the boundary conditions $E^a_j \propto \delta^a_j$ these tangent 
spaces get identified in leading order so that it is not surprising that one can ``undo'' an unwanted 
asymptotic rotation by an internal one. This cannot work in the $U(1)^3$ case because the Gauss constraint does not generate internal rotations.
  
\section{Conclusion}
Motivated by the fact that the $G_N \to 0$ limit of Euclidian general relativity is an interesting toy model for GR, this paper is devoted to study its boundary conditions yielding well-defined symplectic structure and finite and integrable charges associated with the asymptotic symmetries. We have demonstrated that in $U(1)^3$ model, the boundary terms spoiling functionally differentiability of the constraints are exact one-forms, and therefore all constraints can be improved to differentiable functionals. However, these functionals are not finite for boosts and rotations and we have shown that the reason is the lack of the non-Abelian term, that is $\epsilon_{ijk}A_a^j E^a_k$, in the Gauss constraint of this model compared to that of general 
relativity.

The other motivation for this paper came from our companion paper \cite{B.T}
where we needed the decay behaviour of the phase space variables in order 
to select appropriate gauge fixings and Green functions.\\
\\
\\
{\bf Acknowledgements}\\
\\
S.B. thanks the Ministry of Science, Research and Technology of Iran and 
FAU Erlangen-N\"urnberg for financial support.

\end{document}